\documentclass[sigconf]{acmart}

\usepackage{ctable}
\usepackage{multirow}
\usepackage{balance}
\usepackage{booktabs}
\usepackage{hyphsubst}
\usepackage{graphicx}
\usepackage{amsmath}
\usepackage{url}
\usepackage{hhline}
\usepackage[caption = false]{subfig}

\newcounter{lizcounter}
\DeclareRobustCommand{\liz}[1]{\textbf{/* #1 (liz) */}\stepcounter{lizcounter}\typeout{LaTeX Warning: liz comment \thelizcounter: #1 (line \the\inputlineno)}}

\newcounter{findingscounter}

\newif\ifworkinprogress
\workinprogresstrue

\ifworkinprogress
 \newcommand{\ms}[1]{\textcolor{blue}{[Markus] #1}}
 \newcommand{\ez}[1]{\textcolor{green}{[Eva] #1}}
 \newcommand{\dk}[1]{\textcolor{red}{[Dominik] #1}}
\else
 \newcommand{\ms}[1]{}
 \newcommand{\ez}[1]{}
 \newcommand{\dk}[1]{}
\fi

\newcommand{\para}[1]{\vspace{2mm}\noindent\textbf{#1}}

\acmYear{2019} 
\setcopyright{none}
\acmConference{ESCSS 2019}{September 02-04, 2019}{Zurich, Switzerland}\acmPrice{15.00}\acmDOI{xx.xxx/xxxxxx.xxxxxx}

\begin{document}

\title{Modeling Artist Preferences of Users with Different Music Consumption Patterns for Fair Music Recommendations}

\author{Dominik Kowald}
\authornote{Both authors contributed equally to this work.}
\affiliation{%
  \institution{Know-Center GmbH}
  \city{Graz, Austria} 
}
\email{dkowald@know-center.at}

\author{Elisabeth Lex}
\authornotemark[1]
\affiliation{%
  \institution{Graz University of Technology}
  \city{Graz, Austria} 
}
\email{elisabeth.lex@tugraz.at}

\author{Markus Schedl}
\affiliation{%
  \institution{Johannes Kepler University Linz}
  \city{Linz, Austria} 
}
\email{markus.schedl@jku.at}

\begin{abstract}
Music recommender systems have become central parts of popular streaming platforms such as Last.fm, Pandora, or Spotify to help users find music that fits their preferences. These systems learn from the past listening events of users to recommend music a user will likely listen to in the future. Here, current algorithms typically employ collaborative filtering (CF) utilizing similarities between users' listening behaviors. Some approaches also combine CF with content features into hybrid recommender systems~\cite{celma:springer:2010}. 

\para{Problem \& objective.} While music recommender systems can provide quality recommendations to listeners of mainstream music artists, recent research~\cite{schedl_bauer:jmm:2018,Oord2013:DCM} has shown that they tend to discriminate listeners of unorthodox, low-mainstream artists. This is foremost due to the scarcity of usage data of low-mainstream music as music consumption patterns are biased towards popular artists~\cite{Oord2013:DCM,celma:springer:2010}. 

Thus, the objective of our work is to provide a novel approach for modeling artist preferences of users with different music consumption patterns and listening habits. We focus on three user groups: (i) LowMS (i.e., listeners of unorthodox, niche music), (ii) HighMS (i.e., listeners of mainstream music), and (iii) MedMS (i.e., listeners of music that lies in between). The main problem we address in this work is how to exploit variations in listening habits to avoid discrimination of users, whose listening behavior differs significantly from the mainstream. With that, we aim to realize fair music recommendations in the sense that recommendations are not biased towards the mainstream. 

\para{Approach \& method.} In our work, we model user listening behavior on the level of music artists to describe a user's music taste. Since a user's music artist preferences may change over time~\cite{Park2010}, we take temporal drifts of a user's music listening habits into consideration.

To do so, we utilize the Base-Level Learning (BLL) equation from the cognitive architecture ACT-R~\cite{anderson2004integrated} to model music listening habits. The BLL equation accounts for the time-dependent decay of item exposure in human memory. It quantifies the usefulness of a piece of information based on how frequently and how recently it was accessed by a user and models this time-dependent decay using a power-law distribution. We have utilized the BLL equation in our previous works to recommend tags in social bookmarking systems~\cite{Kowald2016bllhypertext} and to recommend hashtags in Twitter~\cite{www_hashtag_2017}.

In the present paper, we build upon these results, and we adopt the BLL equation to model the listening habits of users in our three groups and predict their music artist preferences. We name our approach $BLL_u$ and demonstrate the efficacy of $BLL_u$ using the \textit{LFM-1b} dataset~\cite{schedl2016lfm}, which contains listening histories of more than 120,000 Last.fm users, amounting to 1.1 billion individual listening events over nine years\footnote{The dataset is freely available via \url{http://www.cp.jku.at/datasets/LFM-1b/}}.

Additionally, the dataset contains demographic data such as age and gender as well as a ``mainstreaminess'' factor~\cite{schedl2015tailoring}, which relates a user's artist preferences to the aggregated preferences of all users (i.e., the mainstream). 
Based on this factor, we assign the users in our dataset to one of the three groups: (i) LowMS, (ii) MedMS, and (iii) HighMS. Thus, the 1000 users with the lowest mainstreaminess are in the LowMS group, the 1000 users with a mainstreaminess value centered around the median are in the MedMS group, and the users with the highest values are in the HighMS group. We summarize the dataset statistics of these groups in Table~\ref{tab:datasets} and evaluate our proposed $BLL_u$ approach for all three user groups.

\begin{table}[t]
  \setlength{\tabcolsep}{7pt} 
  \centering
    \begin{tabular}{l||ccccc}
    \specialrule{.2em}{.1em}{.1em}
  Group  & $|U|$   & $|A|$  & $|LE|$  & $|A/U|$   & $|MS|$   \\\hline 
  LowMS  & 1,000  & 82,417 & 6,915,352 & 239 & .125  \\\hline
  MedMS  & 1,000    & 86,249  & 7,900,726 & 496 & .379 \\\hline
  HighMS & 1,000 & 92,690 & 8,251,022 & 1,194 & .688  \\
  \specialrule{.2em}{.1em}{.1em}        
    \end{tabular}
    \caption{Dataset statistics of the LowMS, MedMS, and HighMS Last.fm user groups. Here, $|U|$ is the number of distinct users, $|A|$ is the number of distinct artists, $|LE|$ is the number of listening events, $|A/U|$ is the average number of artists listened by a user and  $|MS|$ is the average mainstreaminess value per user group.\vspace{-7mm}}
  \label{tab:datasets}
\end{table}

\para{Contributions \& results.} The contributions of our work are two-fold. Firstly, we propose our $BLL_u$ approach
that is designed to model and predict artist preferences to provide personalized, fair 
music recommendations. As our notion of fairness is related to popularity bias (i.e., of the mainstream), we model the user's preference for an artist by considering how often this individual user has listened to this artist.
 Additionally, since music preferences are dynamic, we incorporate the user's temporal drifts of artist preferences into our model.  
Secondly, we evaluate our approach on three different groups of Last.fm users based on the distance of their listening behavior to the mainstream: (i) LowMS, (ii) MedMS, and (iii) HighMS.

For our evaluation, we follow good practice in the field of information retrieval and recommender systems by splitting our user groups into train and test sets. We employ a time-based split, and we put the 1\% most recent listening events of each user into the test set and keep the remaining listening events for training. For our three user groups, this procedure leads to three test sets with 68,651 listening events for LowMS, 78,511 listening events for MedMS, and 82,030 listening events for HighMS. For each user, we aim to predict the artists in these listening events\footnote{Our evaluation framework is freely available via \url{https://github.com/learning-layers/TagRec}}.

Figure~\ref{fig:results_plots} illustrates the results of our evaluation in form of recall/precision plots. Here, we compare $BLL_u$ to four baselines: (i) $TOP$, which recommends the most popular artists of all users, (ii) $CF_u$, which recommends artists using collaborative filtering, (iii) $POP_u$, which recommends the most popular artists of a specific user, and (iv) $TIME_u$, which recommends the artists a particular user has listened to most recently. We find that for all groups, $BLL_u$ leads to the best accuracy results for predicting music artists and provides especially good results for the LowMS group. Interestingly, we also find that the time-based approach $TIME_u$ provides even better accuracy results than $BLL_u$ when only predicting 1 or 2 artists.

\begin{figure}[t]
   \centering
  \captionsetup[subfigure]{justification=centering}
  \subfloat[User group: LowMS]{ 
      \includegraphics[width=0.50\textwidth]{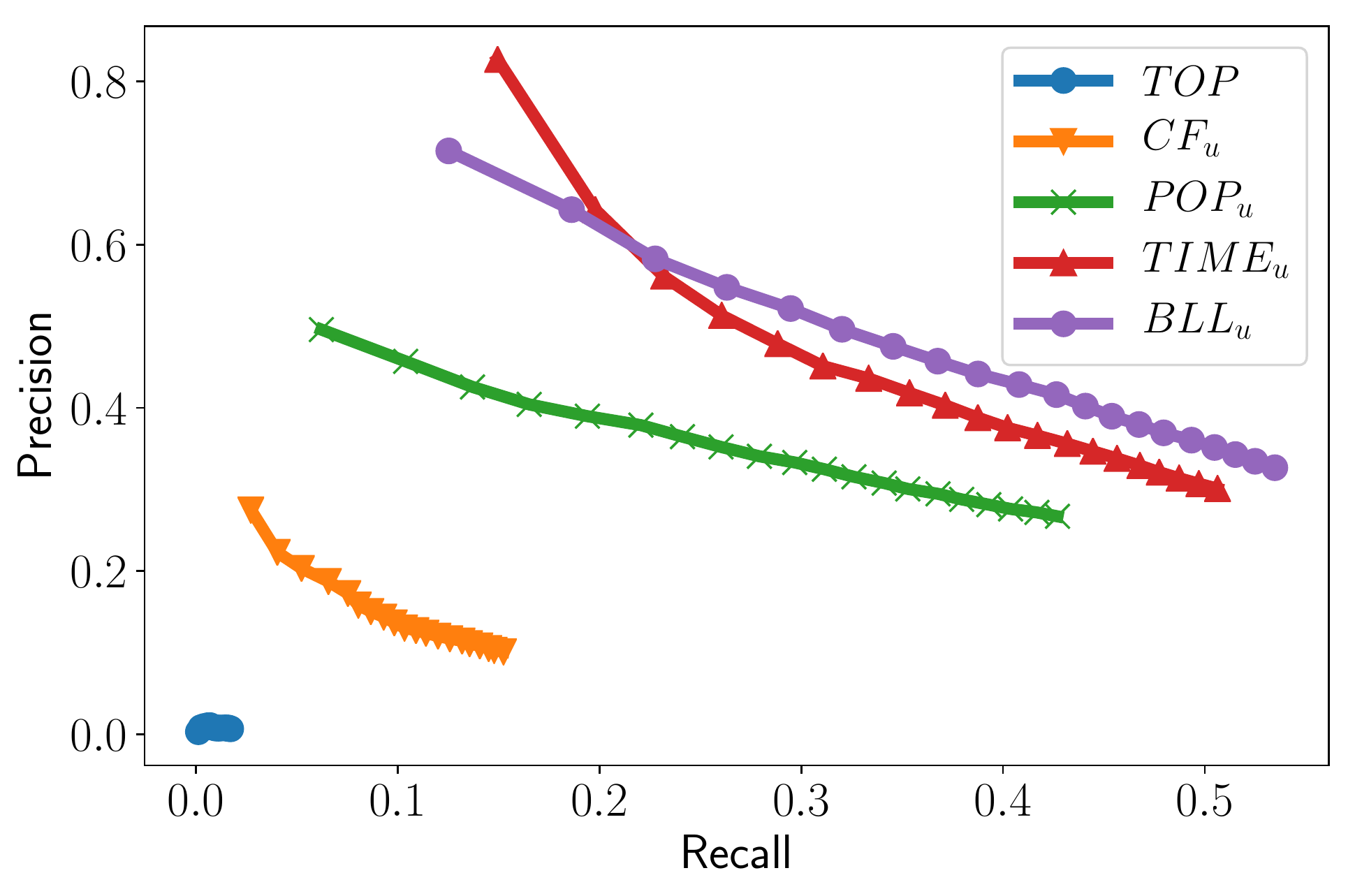}
   }\\
  \subfloat[User group: MedMS]{ 
      \includegraphics[width=0.50\textwidth]{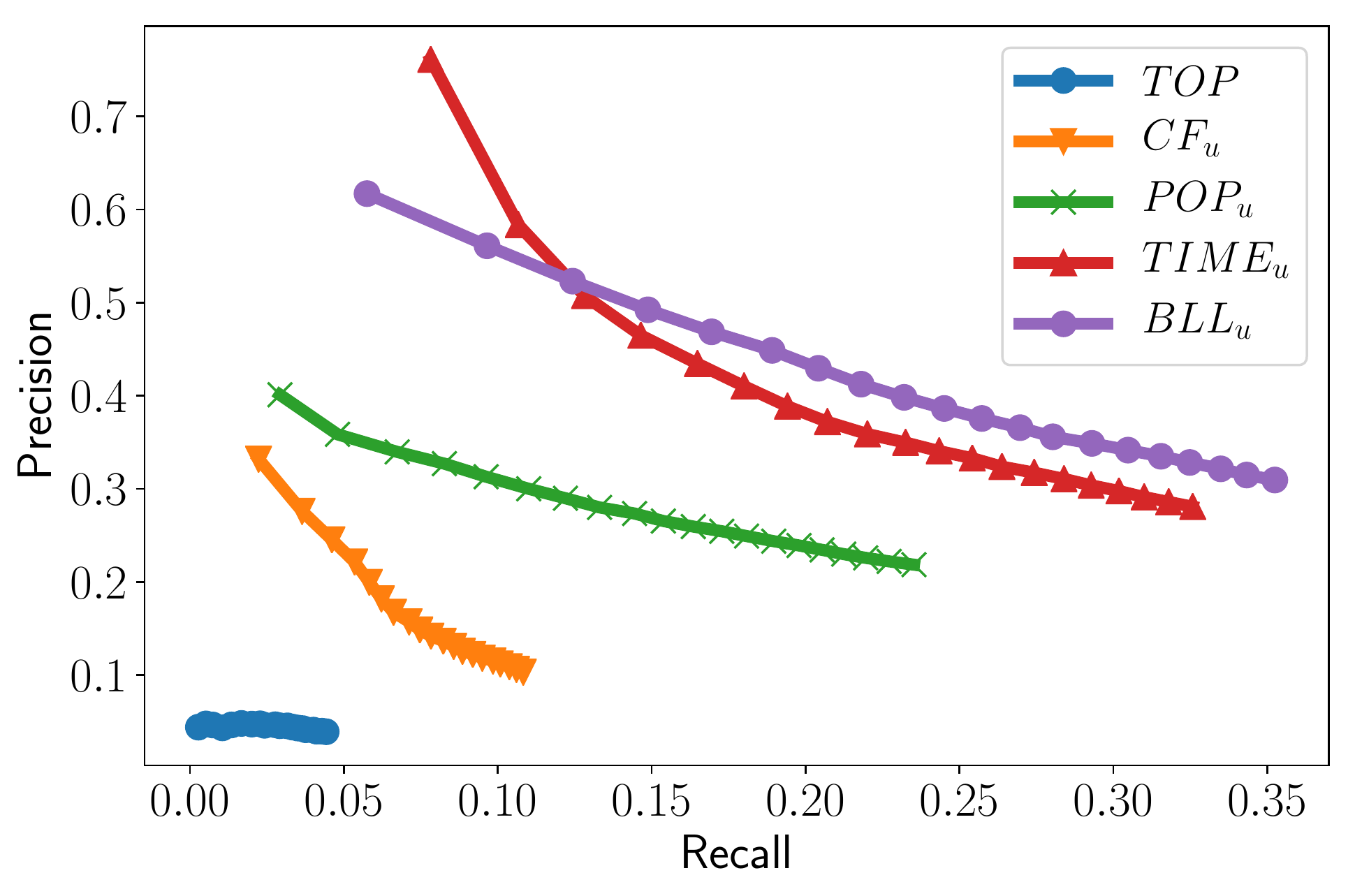} 
   }\\
  \subfloat[User group: HighMS]{ 
      \includegraphics[width=0.50\textwidth]{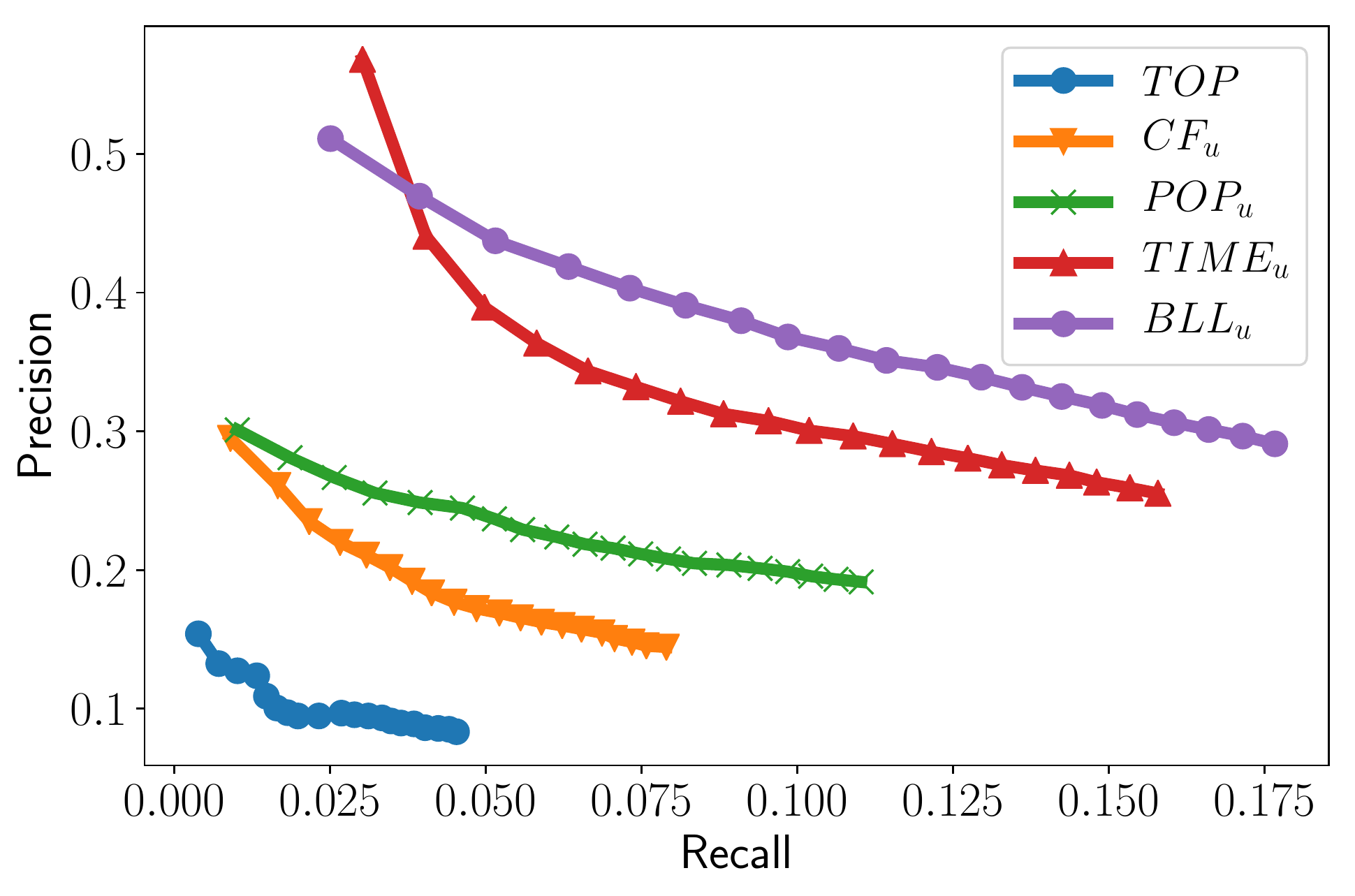} 
   }
   \caption{Recall/precision plots of the baselines and our $BLL_u$ approach for the three user groups LowMS, MedMS, and HighMS, and for $k = 1 \ldots 20$ predicted artists. We see that $BLL_u$ provides the best prediction accuracy results for all user groups.\vspace{-3mm}}
  \label{fig:results_plots}
\end{figure}

\para{Future work.}
We plan to extend our analysis to include more sophisticated mainstreaminess measures based on rank-order correlation or Kullback-Leibler divergence~\cite{schedl_bauer:jmm:2018} since our current mainstreaminess measure is rather simplistic. Furthermore, we aim to integrate our findings into fair music recommendation algorithms (e.g., for songs), with particular attention to avoid discrimination of the low mainstreaminess group, since standard collaborative filtering approaches do not provide suitable music recommendations for this user group~\cite{schedl2015tailoring}. For example, we plan to integrate the BLL-preference values we obtain for a specific user and a particular artist via our approach as a context dimension into a matrix factorization-based approach such as the one presented in~\cite{koenigstein2011yahoo}.

\para{Keywords.} User Groups; Music Preference Prediction; Fair Music Recommendations; Discrimination; Time-Aware Recommendations

\end{abstract}


\maketitle






\bibliographystyle{ACM-Reference-Format}

\end{document}